# Physical properties of RhGe and CoGe single crystals synthesized under high pressure

Shangjie Tian[1,2,3,#], Xiangjiang Dong[4,#], Bowen Zhang[4], Zhijun Tu[2,3], Runze Yu[4,*], Hechang Lei[2,3,*], and Shouguo Wang[1,*]

[1]*Anhui Provincial Key Laboratory of Magnetic Functional Materials and Devices, School of Materials Science and Engineering, Anhui University, Hefei 230601, China*

[2]*School of Physics and Beijing Key Laboratory of Optoelectronic Functional Materials & MicroNano Devices, Renmin University of China, Beijing 100872, China*

[3]*Key Laboratory of Quantum State Construction and Manipulation (Ministry of Education), Renmin University of China, Beijing 100872, China*

[4]*Center for High-Pressure Science and Technology Advanced Research, Beijing 100093, China*

[#]These authors contributed to this work equally.

[*]Corresponding authors: R.Z.Y. (runze.yu@hpstar.ac.cn); H.C.L (hlei@ruc.edu.cn); S.G.W. (sgwang@ahu.edu.cn).

**Abstract**

Chiral topological semimetals hosting multifold fermions and exotic surface states represent a frontier in topological materials research. Among them, noncentrosymmetric cubic B20 compounds—notably transition-metal silicides and germanides—offer a unique platform for realizing symmetry-protected topological phases and unconventional optoelectronic responses. Here, we report the physical properties of RhGe and CoGe single crystals with B20 structure in detail. Transport measurements reveal metallic behavior with characteristic Fermi-liquid scaling at low temperatures, while magnetization results confirm paramagnetism in both compounds. In addition, both of materials exhibit low carrier concentrations with small electronic specific heat coefficient, indicating their semimetal feature with weak electronic correlations. Such high-quality CoGe and RhGe single crystals provide a material platform to explore the evolution of multifold fermions and the instability of helicoid-arc surface states with spin-orbit coupling and surface environment in B20 material systems.

## Introduction

The existence of nontrivial band topology generates exotic quantum phenomena in condensed matter physics, establishing topological materials as a major research frontier since the discovery of topological insulators and semimetals [1-3]. Fundamental fermionic particles predicted in high-energy physics—including Dirac, Weyl, and Majorana fermions—emerge as low-energy quasiparticle excitations within these materials. Dirac semimetals host four-fold degenerate nodes exhibiting linear dispersion in all momentum directions, protected by combined inversion, time-reversal, and lattice symmetries [4-7]. Breaking either inversion or time-reversal symmetry splits Dirac nodes into Weyl points (characterized by a Chern number $|C| = 1$), generating distinctive Fermi arc surface states [8-11]. Chirality, degeneracy, and dimensionality further define diverse topological phases [12-14], while non-symmorphic symmetries enforce high-symmetry-point band crossings that produce higher-fold fermions. Recent advances reveal massless fermions beyond standard model analogues—such as spin-1 chiral fermions, double Weyl fermions, and spin-3/2 Rarita-Schwinger-Weyl (RSW) fermions—characterized by higher quantized Chern numbers [15-22].

Transition metal monosilicides and monogermanides crystallizing in the cubic noncentrosymmetric B20 structure (space group $P2_13$, No. 198) exhibit a rich spectrum of magnetic, topological, and transport properties [15-19, 23-40]. The inherent structural chirality in these materials dictates both electronic and magnetic behaviors, leading to some unusual phenomena in real and reciprocal spaces. On the one hand, structural chirality induces a finite spin-orbit-driven Dzyaloshinskii-Moriya interaction, which competes with the Heisenberg exchange coupling. This competition underpins unusual magnetic properties, including the observation of skyrmion phases in compounds such as MnSi, FeGe, MnGe, and solid solutions like $Fe_{1-x}Co_xSi$ and $Fe_{1-x}Mn_xSi$ [24, 25]. On the other hand, in nonmagnetic B20 systems, structural chirality enables chiral topological semimetal states. These states host non-symmorphic symmetry-protected multifold fermions at high-symmetry points [22, 26-28]. For example, when neglecting spin-orbit coupling (SOC), the band structures of CoSi exhibit a threefold degenerate crossing (spin-1 fermion) at the $\Gamma$ point and a four-fold degenerate crossing (double Weyl fermion) at the $R$ point, carrying chiral charges of +2 and -2, respectively [22, 26-28]. Such unconventional quasiparticles can give rise to novel optoelectronic responses, such as the circular photogalvanic effect [16, 29]. The surface states of CoSi and RhSi also demonstrate long Fermi arcs but they possess two different types of helicoid arc van Hove singularities [30, 31], indicating the diversity of surface states (SSs) due to distinct chemical environments of the surface and SOC. Furthermore, theoretical studies have recently substantiated the potential for topological superconductivity in B20-structured RhGe [32].

Despite their intriguing topological properties, the studies on B20 RhGe and CoGe, which are isoelectronic to RhSi and CoSi, remain scarce and controversial due to the significant challenges associated with their single-crystal synthesis, which necessitates high-pressure and high-temperature conditions [33-36]. Previous study of RhGe polycrystal suggested there is a coexistence of weak itinerant ferromagnetism ($T_C \sim 140$ K) and superconductivity ($T_c \sim 4.3$ K) in this material [35], but our preliminary experimental results on RhGe single crystals indicate that both behaviors are absent [37]. In contrast, CoGe exhibits a Pauli paramagnetism [38, 39]. In this work, we study the physical properties of CoGe and RhGe single crystals grown by using high-pressure and high-temperature method. Both materials exhibit paramagnetic metallic behaviors with low concentrations of electron-type charge carriers.

## Methods

Single crystals of RhGe and CoGe were grown under high pressure in a cubic anvil apparatus. High-purity elemental precursors (Rh: 99.99 %, Co: 99.95 % and Ge: 99.99 %) were homogenized by grinding with stoichiometric ratios of Rh : Ge = 1 : 1.2 (excess Ge serving as solvent flux) and Co : Ge = 1 : 1, respectively, sealed in hBN capsules (Au and Pt capsules were avoided because they can react with Ge under high temperature and high pressure). They were then positioned at the center of the high-pressure cell and compressed to 5 GPa. For RhGe (CoGe), the temperature was ramped to 1250 °C (1100 °C) at 10 °C/min, maintained for 3 h (2 h), slowly cooled to 800 °C (700 °C), and finally quenched to ambient temperature prior to decompression.

The powder X-ray diffraction (PXRD) patterns of crushed RhGe and CoGe crystals were performed using a high-resolution diffractometer (Bruker D8) using Cu $K_\alpha$ radiation ($\lambda$ = 1.5406 Å) at room temperature. The single-crystal orientations of RhGe and CoGe were determined using a Laue in backscattering mode (SmartLab, Rigaku). Scanning electron microscopy (SEM) imaging and energy-dispersive X-ray (EDX) spectroscopy for chemical composition analysis were performed using a JEOL JSM-7000F field-emission scanning electron microscope. Electrical transport measurements were conducted using a superconducting magnetic system (Cryomagnetics, C-Mag Vari-9). Magnetization and heat capacity measurements were performed using a Quantum Design magnetic property measurement system (MPMS3) and physical property measurement system (PPMS-14T), respectively.

## Results and Discussions

The RhGe and CoGe compounds crystallize in the non-centrosymmetric cubic chiral B20 structure (space group $P2_13$, No. 198) [23, 33-35, 39], isostructural to CoSi [26-28]. This non-symmorphic chiral framework features a distorted rocksalt-type arrangement with atomic displacements along the [111] direction (Fig. 1(a)). Figures 1(b) and 1(c) displays the PXRD patterns for both compounds measured at room temperature and ambient pressure, accompanied by Rietveld refinement profiles, confirming the phase-pure B20 structures. The refinement parameters ($R_{wR}$ = 7.504 % for RhGe and 1.046 % for CoGe) demonstrate high model reliability. The fitted $a$-axial lattice parameter is 0.4859(2) nm and 0.4640(1) nm for RhGe and CoGe, respectively, which are in agreement with the reported values in literature for polycrystalline samples [33, 35, 38-40]. The larger lattice constant of RhGe than that of CoGe can be ascribed to the larger ionic radius of Rh than Co. The B20 structure contains 8 atoms per unit cell with Rh/Co and Ge occupying Wyckoff positions 4$a$ at fractional coordinates: ($u$, $u$, $u$), ($u$ + 0.5, 0.5 - $u$, -$u$), ($u$, 0.5 + $u$, 0.5 - $u$), and (0.5 - $u$, $u$, 0.5 + $u$). The refined positional parameter $u$ is 0.1086(2) for Rh and 0.9023(4) for Ge in RhGe, while CoGe shows $u$ = 0.135(4) for Co and 0.837(2) for Ge. EDX analysis determined the composition to be Rh : Ge = 1.08(1) : 1 for RhGe, which is close to the ideal stoichiometric ratio. For CoGe, however, the measured ratio of Co : Ge was 1.25(1) : 1 suggests the presence of excess Co and related defects. Figures 1(d) and 1(e) present the Laue diffraction patterns of RhGe and CoGe single crystals along the high-symmetry (100) plane, exhibiting sharp, well-indexed diffraction spots with cubic symmetry, further attesting to the high quality and structural integrity of grown single crystals.

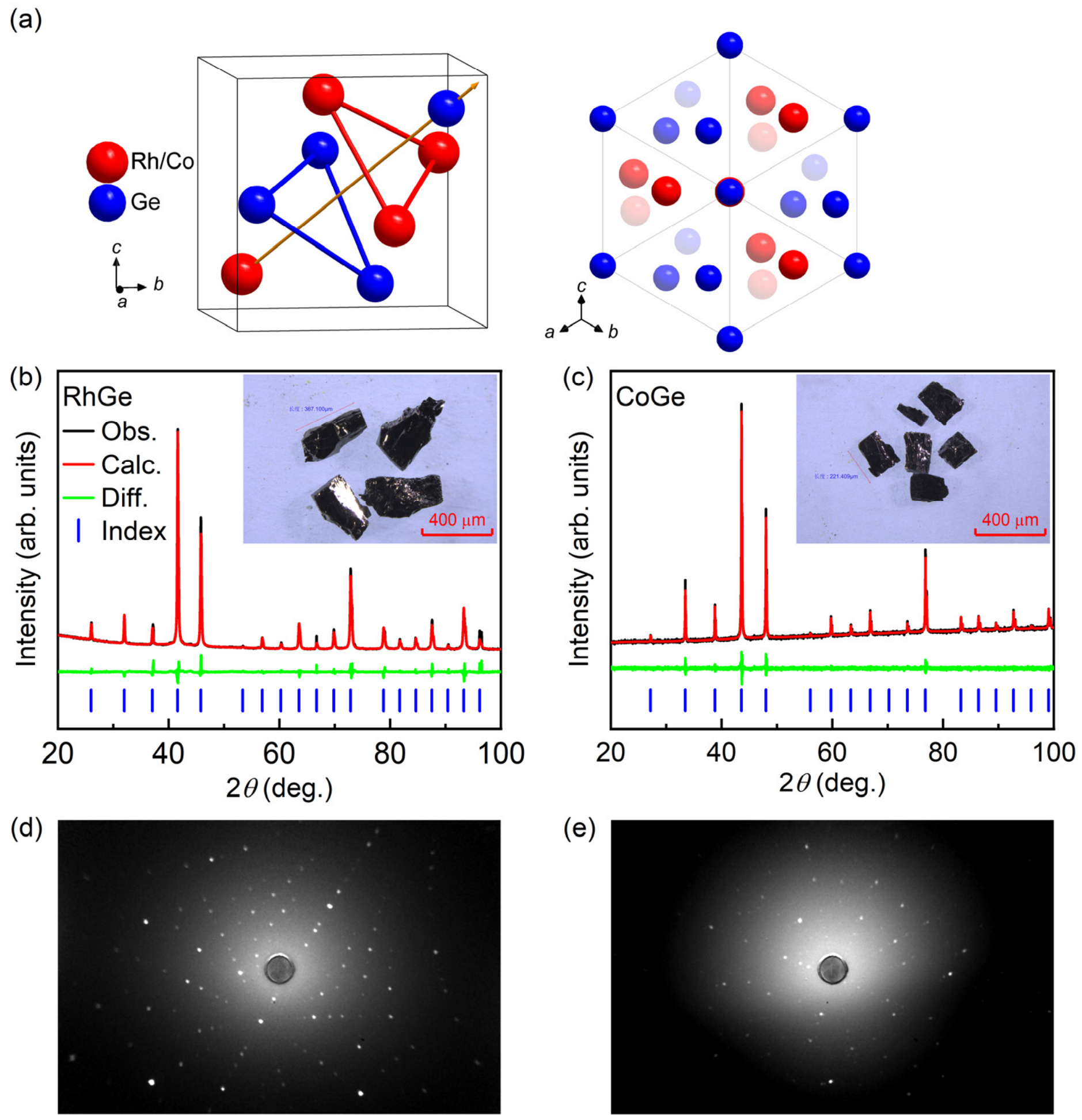

**Figure 1.** (a) The B20 cubic crystal structure of RhGe and CoGe with origin shift (0.23, 0.23, 0.23). The blue and red balls represent Rh/Co and Ge atoms, respectively. The orange arrow in the left panel indexes the [111] direction. The helix arrangements are clearly visible in the top view along the [111]-direction (right panel). PXRD patterns of (b) RhGe and (c) CoGe along with the Rietveld refinements. The insets show the photograph of each single crystal. Laue diffraction patterns of (d) RhGe and (e) CoGe single crystal.

The temperature-dependent electrical resistivity $\rho_{xx}(T)$ of RhGe and CoGe single crystals are shown in Figs. 2(a) and 2(b). Both of them exhibit metallic behaviors in general and the residual resistivity ratios *RRR* is about 27 and 2 for RhGe and CoGe, respectively. For RhGe, $\rho_{xx}(T)$ decreases linearly from ~ 150 μΩ cm at 300 K down to ~ 70 K, indicating the dominant electron-phonon scattering at the high-temperature region. The $\rho_{xx}(T)$ can be fitted using $\rho_{xx}(T) = \rho_0 + AT^n$ between 20 - 70 K, the obtained $n$ is 2.05(2), implying that the Fermi-liquid behavior appears at low-temperature region in RhGe. It is noted that superconductivity is absent in present RhGe crystals when temperature above 1.8 K, consistent with our previous results [37]. As emphasized in Ref. [35], superconductivity in RhGe is highly sensitive to sample composition and applied pressure. The EDX analysis revealed a slight compositional difference between our single crystals and the superconducting polycrystals. We therefore suggest that the absence of superconductivity in our samples may be attributed to these subtle differences in composition and/or residual strain, which can inherently arise from differences in synthesis methods. For CoGe crystal, it displays similar trends to RhGe but with lower resistivity (~ 120 μΩ cm at 300 K). The $\rho_{xx}(T)$ curve shows a linear behavior with lowering $T$ down to 180 K and changes to a $T^{2.06(4)}$ dependence between 40 - 70 K, consistent with Fermi-liquid behavior. When $T < 25$ K, a subtle resistivity upturn can be observed. Such low-$T$ resistivity increase are generally attributed to three mechanisms: weak localization (WL), Kondo effect and electron-electron interaction (EEI) *etc*. However, WL and Kondo effect will lead to a negative MR when applying a magnetic field and are safely excluded here due to being inconsistent with the positive MR of CoGe (discuss below) [38-40]. As depicted in the inset of Fig 2(b), the upturn part in conductivity curve follows a $T^{1/2}$ dependence, suggest the existence of EEI in the presence of disorder [38]. We propose that the disorder may arise from the excess Co defects in CoGe.

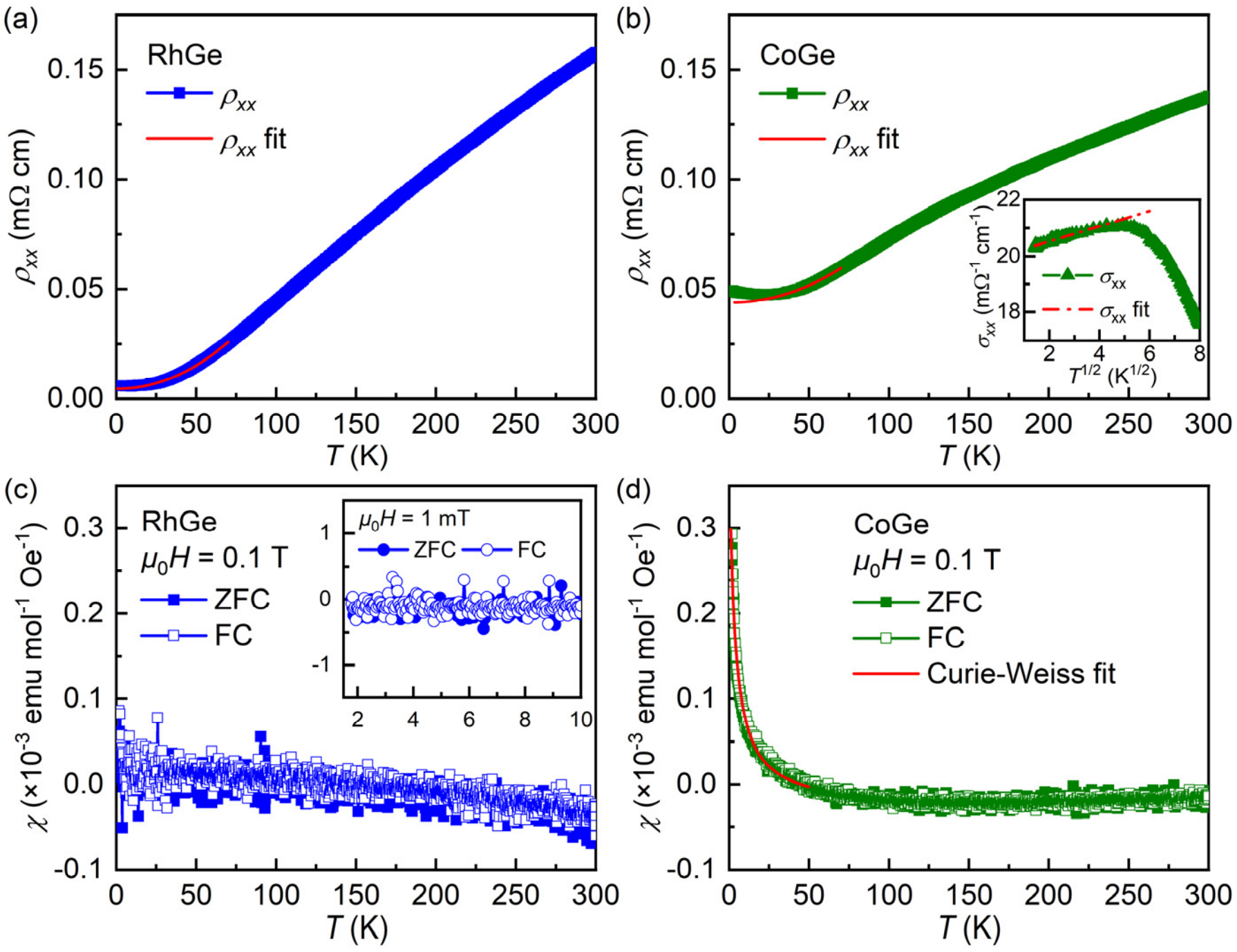

**Figure 2.** Temperature dependence of $\rho_{xx}(T)$ of (a) RhGe and (b) CoGe single crystals. The red solid lines show the fits using the formula $\rho_{xx}(T) = \rho_0 + AT^n$. Inset of (b) shows longitudinal conductivity $\sigma_{xx}$ as a function of $T^{1/2}$ of CoGe single crystals, accompanied by a linear fit. Temperature dependence of $\chi(T)$ at field of 0.1 T with ZFC and FC modes for (a) RhGe and (b) CoGe single crystal, respectively. Inset of (a) shows $\chi(T)$ data at $\mu_0H = 1$ mT with ZFC mode. The red lines in (b) show the fits using the modified Curie-Weiss law $\chi = \chi_0 + C/(T - T_\theta)$.

Temperature dependence of magnetic susceptibility $\chi(T)$ of RhGe and CoGe at $\mu_0H = 0.1$ T with zero-field-cooling (ZFC) and field-cooling (FC) modes are shown in Figs. 2(c) and 2(d). For RhGe, the $\chi$(T) curve exhibits a weak temperature dependence in the whole temperature range (Fig. 2(c)), suggesting a Pauli paramagnetic behavior. At low-temperature region, the absence of diamagnetic signal in $\chi(T)$ curve at $\mu_0H = 1$ mT with ZFC mode (inset of Fig. 2(b)) confirms there is no superconducting transition in present RhGe single crystals, consistent with resistivity data. For CoGe, the $\chi(T)$ curve shows a similar behavior to RhGe at $T > 50$ K (Fig. 2(d)), implying the same Pauli paramagnetism. It is consistent with prior report on CoGe polycrystal [41, 42]. When $T < 50$ K, however, the $\chi(T)$ curve of CoGe exhibits a slight upturn behavior (Fig. 2(d)), likely due to trace magnetic impurities, spin fluctuations or weak electronic correlations. The fitted Curie constant is 0.0014(3) emu K mol$^{-1}$ Oe$^{-1}$ using the modified Curie-Weiss law ($\chi = \chi_0 + C/(T - T_\theta)$), corresponding to 0.11(2) $\mu_B$/f.u.. The observed excess Co from EDX analysis suggests that these atoms may act as magnetic impurities. If assuming these magnetic impurities originates from Co element with magnetic moment 3.87 $\mu_B$, the estimated molar content of magnetic impurities are below 2.8(6) %.

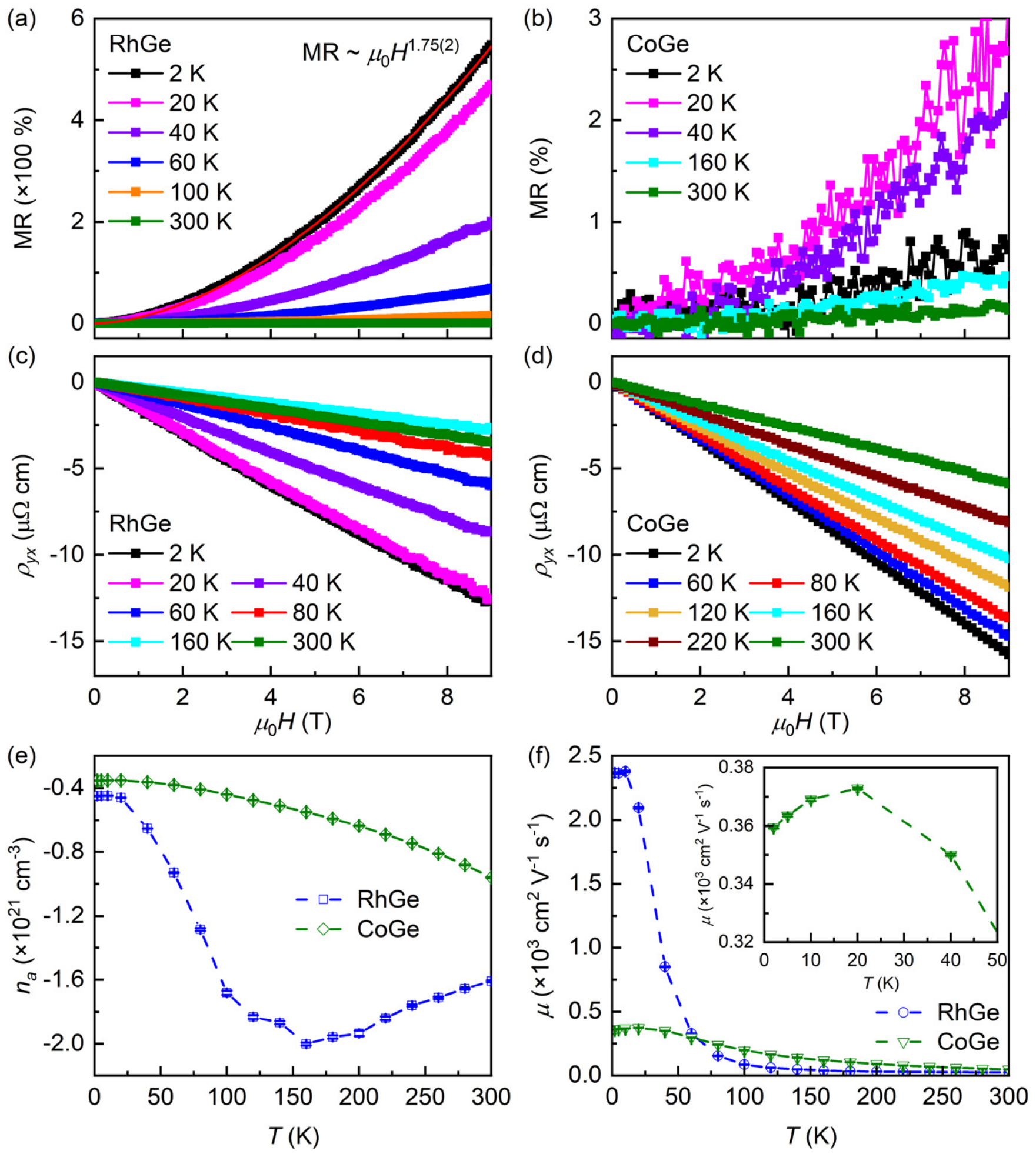

**Figure 3.** Field dependence of MR up to $\mu_0H = 9$ T at various temperatures for (a) RhGe and (b) CoGe single crystal, respectively. The color codes in (a) and (b) are the same. The red solid line in (a) is the fit of MR data at 2 K using MR = $A(\mu_0H)^n$. (c) and (d) The $\rho_{yx}(\mu_0H)$ of RhGe and CoGe as a function of $\mu_0H$ between 2 K and 300 K. The color codes in (c) and (d) are the same. (e) Temperature dependences of $n_a(T)$ for RhGe and CoGe derived from the linear fits of $\rho_{yx}(\mu_0H)$ curves at high-field region (4 – 9 T). (f) Derived temperature dependences of $\mu(T)$ of RhGe and CoGe. Inset: enlarged part of $\mu(T)$ below 50 K for CoGe.

Figures 3(a) and 3(b) present the magnetoresistance (MR) of RhGe and CoGe at various temperatures. RhGe exhibits a large positive MR (~ 600 %) at 2 K, which diminishes rapidly with increasing temperature. Moreover, the MR curve can be fitted well using the formula MR = $A(\mu_0H)^n$ with $n$ = 1.75(2). Such nearly parabolic field dependence suggests that there may be multiply bands with different carrier type (electron or hole) near Fermi level $E_F$. In contrast, CoGe displays a weak positive MR (< 3 %) across the entire measured range (2 – 300 K, 0 – 9 T), comparable to those of CoGe polycrystal [42] and CoSi grown via the chemical vapor transport method [44]. The Hall resistivity $\rho_{yx}(\mu_0 H)$ as a function of magnetic field at various temperatures is shown in Figs. 3(c) and 3(d) for both materials. All curves exhibit negative slopes, indicating the dominance of electron-type carriers in the transport properties. By linearly fitting the $\rho_{yx}(\mu_0 H)$ curves at high-field region (4 – 9 T) and combining them with zero-field $\rho_{xx}(T)$ data, the temperature dependence of apparent carrier concentration $n_a(T)$ and mobility $\mu(T)$ can be determined based on single-band model ($R_H = \rho_{yx}(\mu_0 H)/\mu_0 H = 1/n_a e$ and $\rho_{xx} = ne\mu$, where $R_H$ is Hall coefficient and $e$ is elementary charge). As shown in Fig. 3(e), the carrier concentrations of both compounds at 2 K are comparable: $n_a \sim 4.5 \times 10^{20}$ cm$^{-3}$ and $3.5 \times 10^{20}$ cm$^{-3}$ for RhGe and CoGe, respectively. Moreover, these values of $n_a$ are also close to that in CoSi [44, 45]. With increasing temperature $T$, the $n_a$ of RhGe increases rapidly, reaching a maximum at $T \sim 160$ K, and then starts to decrease gradually. It implies that at high temperature there may have more thermal activated hole-type carriers mainly from the band at $M$ point in Brillouin zone, which leads to the decrease of $n_a$. In contrast, CoGe shows a monotonic mild increase of $n_a$ with increasing $T$. For RhGe, it exhibits a significantly higher mobility ($\mu \sim 2366$ cm$^2$ V$^{-1}$ s$^{-1}$) than that of CoGe ($\mu \sim 359$ cm$^2$ V$^{-1}$ s$^{-1}$) at 2 K (Fig. 3(f)). At high $T$, The $\mu$ of RhGe undergoes a rapid decay to ~ 350 cm$^2$ V$^{-1}$ s$^{-1}$ at $T = 50$ K and finally deceases to ~ 25 cm$^2$ V$^{-1}$ s$^{-1}$ at 300 K. For CoGe, its $\mu$ peaks at ~ 370 cm$^2$ V$^{-1}$ s$^{-1}$ near 25 K (inset of Fig. 3(f)) before declining to ~ 47 cm$^2$ V$^{-1}$ s$^{-1}$ at 300 K slowly. Such behavior explains the more obvious low-temperature MR of RhGe when compared to CoGe and its rapid suppression with $T$. It also suggests that the upturn feature of $\rho_{xx}(T)$ for CoGe at $T \sim 25$ K (Fig. 1(c)) originates from the non-monotonic behavior of $\mu$ not the evolution of $n_a$.

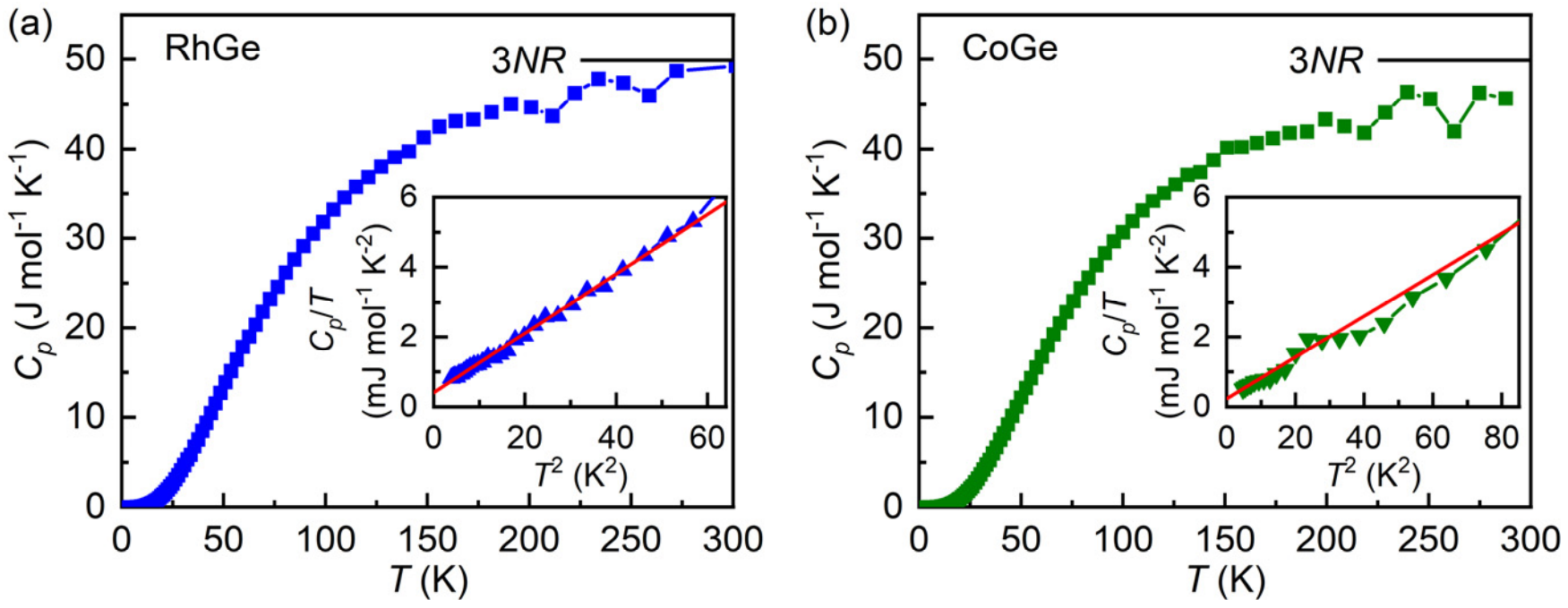

**Figure 4.** Temperature dependence of $C_p(T)$ from 2 K to 300 K at zero field for (a)

RhGe and (b) CoGe, respectively. Insets of (a) and (b) show the relationship between $C_p/T$ and $T^2$. The red solid lines represent the fits using the formula $C_p/T = \gamma + \beta T^2$.

Specific heat measurements of RhGe and CoGe single crystals, shown in Figs. 4(a) and (b), reveal their thermodynamic properties across 2 – 300 K. For RhGe single crystal, no anomaly (e.g., specific heat jump) at specific heat $C_p(T)$ curve is observed at 2.6 K (Fig. 4(a)), consistent with above resistivity and magnetic susceptibility results. It confirms the absence of superconductivity in present RhGe single crystal. The high-temperature $C_p(T)$ for both compounds approaches the Dulong-Petit limit (~ 57 J mol$^{-1}$ K$^{-1}$ at 300 K), consistent with the classical value $3NR$ ($N = 2$ atoms per formula unit, $R = 8.314$ J mol$^{-1}$ K$^{-1}$). At low-temperature region, the $C_p(T)$ data can be fitted well using the formula $C_p/T = \gamma + \beta T^2$, where $\gamma$ represents the electronic specific heat coefficient and $\beta$ the lattice contribution (insets in Figs. 4(a) and (b)). The fitted $\gamma$ of RhGe is 0.40(1) mJ mol$^{-1}$ K$^{-2}$ and $\beta$ = 0.084(9) mJ mol$^{-1}$ K$^{-4}$, yielding a Debye temperature $\Theta_D$ = 357(6) K via $\Theta_D = (12\pi^4 NR/5\beta)^{1/3}$. CoGe shows a comparable value of $\gamma$ with RhGe (~ 0.24(1) mJ mol$^{-1}$ K$^{-2}$) and a larger $\Theta_D$ = 403(4) K, derived from the value of $\beta$ = 0.059(2) mJ mol$^{-1}$ K$^{-4}$. The latter can be explained by the smaller atomic weight of Co than that of Rh. The values of $\gamma$ in both RhGe and CoGe is similar to that of CoSi ($\gamma$ ~ 0.5 – 1.2 mJ mol$^{-1}$ K$^{-2}$) [46], but much smaller than MnGe ($\gamma$ = 16 - 20 mJ mol$^{-1}$ K$^{-2}$) [41, 42] and FeGe ($\gamma$ = 9 mJ mol$^{-1}$ K$^{-2}$) [42] (summarized in Table 1 with other related B20 materials). It demonstrate that the electron correlations in nonmagnetic RhGe, CoGe and CoSi are weaker than those of magnetic B20 systems.

A comparison of the physical parameters of RhGe and CoGe with other silicide and germanide materials with B20 structure is presented in Table 1. Theoretical calculations indicate that RhGe and CoGe are nonmagnetic, and they exhibit remarkably similar electronic band structures, well-described within the rigid band approximation framework. States within 5 eV of the Fermi energy ($E_F$) originate predominantly from transition metal *d*-orbitals, which are understood to govern many of the intriguing properties in these compounds. Specifically, the Fermi surfaces of CoGe and RhGe comprise electron pockets centered at the Brillouin zone center ($\Gamma$ point) and corner ($R$ point), alongside a hole pocket at the $M$ point. Additional, very small hole pockets may exist along the $\Gamma$-$R$ direction. All the Fermi surface pockets are rather small and resemble those of semimetals [22, 42, 47-50]. Our magnetic and electrical transport measurements corroborate the theoretical predictions, confirming that both RhGe and CoGe single crystals are paramagnetic metals with low charge carrier density, similar to CoSi. These features likely arises from a significantly reduced density of states (DOS) at $E_F$ when compared to itinerant ferromagnets like FeGe or MnGe. More importantly, because CoGe and RhGe are isoelectronic to CoSi and RhSi, they should exhibit similar band topology, especially long Fermi arcs. But different strength of SOC and subtle surface environment could lead to different connectivity and instability of SSs. Thus, RhGe and CoGe provide a novel platform to study surface-dominant correlation effects in topological semimetals.

Table 1. Summary of physical parameters of some silicide and germanide materials with B20 structure.

| | RRR | MR (%) | $n$ ($10^{20}$ cm$^{-3}$) | $\mu$ (cm$^2$ V$^{-1}$ s$^{-1}$) | $\gamma$ (mJ mol$^{-1}$ K$^{-2}$) | Properties* | Ref. |
|---|---|---|---|---|---|---|---|
| CoSi | 1 – 30 | -1 – 400 | 1 – 3 | 200 – 7300 | 0.5-1.2 | PM，UCF | [26, 28, 41, 43, 51] |
| CoGe | 2<br>1.8 | <3<br><0.1 | 2.5<br>6 | 359 | 0.24<br>0 – 3 | PM<br>PM, UCF | This work<br>[38-40, 46] |
| RhSi | | | | | | PM, UCF | [15, 16] |
| RhGe | 27<br>10 | 600 | 4.5 | 2366 | 0.40(1)<br>2.9 | PM<br>SC ($T_c$ = 4 K), UCF | This work<br>[32，35] |
| RhSn | 24 | 450 | 4.6 | | | PM, UCF | [19, 52] |
| FeGe | 14 | -0.5 | 2/3 carrier/f.u. | | 9 | HM ($T_C$ = 280 K), Sky | [25, 39] |
| MnGe | 9.3 | <5 | 1/2 carrier/f.u. | | 16 - 20 | HM ($T_C$ = 275 K), Sky | [25, 38, 39] |

* PM = paramagnetic, UCF = unconventional chiral fermions, SC = superconductivity, HM = helimagnetic, Sky= Skyrmion.

## Conclusion

In summary, we successfully grew the high-quality RhGe and CoGe single crystals up to millimeter size under high pressure and high temperature. Transport measurement reveals the metallic behaviors of RhGe and CoGe single crystals at low temperature, while magnetic measurements indicate that both RhGe and CoGe are paramagnets. Neither superconductivity nor ferromagnetism are observed down to 2 K. These high-quality RhGe and CoGe single crystals are prerequisite for further investigation of their novel topological properties using angle-resolved photoemission spectroscopy (APRES). Moreover, this study highlights the ongoing challenges of discovering superconductivity in B20-structured materials.

## Acknowledgments

This work was supported by National Key R&D Program of China (Grants No. 2022YFA1403800 and No. 2023YFA1406500), National Natural Science Foundation of China (Grants No. 12274459, 12474002 and 22171283), and China Postdoctoral Science Foundation (Grant Nos. 2023M730011). This work was supported by the Synergetic Extreme Condition User Facility (SECUF, https://cstr.cn/31123.02.SECUF)